\documentclass[useAMS,usenatbib]{mn2e}

\usepackage{epsfig}
\usepackage{amsmath}
\usepackage{amssymb}
\usepackage{natbib}
\usepackage{threeparttable} 

\usepackage{epstopdf}

\newcommand{\chan}{\textit{Chandra}}
\newcommand{\swift}{\textit{Swift}}
\newcommand{\rxte}{\textit{RXTE}}

\newcommand{\inte}{\textit{Integral}}
\newcommand{\exosat}{\textit{EXOSAT}}
\newcommand{\rosat}{\textit{ROSAT}}

\newcommand{\Msun}{\mathrm{M}_{\odot}}
\newcommand{\lum}{\mathrm{erg~s}^{-1}}
\newcommand{\flux}{\mathrm{erg~cm}^{-2}~\mathrm{s}^{-1}}
\newcommand{\cnts}{\mathrm{counts~s}^{-1}}
\newcommand{\mdot}{\mathrm{M_{\odot}~yr}^{-1}}

\newcommand{\source}{CXOGClb J174804.8--244648}
\newcommand{\exo}{EXO 1745--248}
\newcommand{\intename}{IGR J174880--2446}
\newcommand{\hakuchoname}{XB 1745--25}
\newcommand{\sax}{SAX J1808.4--3658}
\newcommand{\ngcbron}{NGC 6440 X-1}
\newcommand{\xtepulsar}{XTE J0929--314}
\newcommand{\igrpulsar}{IGR 00291+5934}

\def \mnras {MNRAS}
\def \apj {ApJ}

\def \apjl {ApJL}
\def \aap {A\&A}

\def \araa {ARAA}

\def \iaucirc {IAU Circ.}

\title[The Terzan 5 X-ray pulsar in quiescence]{The soft quiescent spectrum of the transiently accreting 11~Hz X-ray pulsar in the globular cluster Terzan~5}
\author[N. Degenaar \& R. Wijnands]
{N. Degenaar\thanks{e-mail: degenaar@uva.nl} \& R. Wijnands\\
Astronomical Institute "Anton Pannekoek", 
University of Amsterdam, 
Postbus 94249, 1090 GE Amsterdam, the Netherlands
\vspace{-0.3cm}
}

\begin{document}

\date{Accepted 2010 December 23. Received 2010 December 23; in original form 2010 November 26
}

\pagerange{\pageref{firstpage}--\pageref{lastpage}} \pubyear{0000}

\maketitle

\label{firstpage}

\begin{abstract} 
We report on the quiescent X-ray properties of the recently discovered transiently accreting 11 Hz X-ray pulsar in the globular cluster Terzan 5. Using two archival \chan\ observations, we demonstrate that the quiescent spectrum of this neutron star low-mass X-ray binary is soft and can be fit to a neutron star atmosphere model with a temperature of $kT^{\infty} \sim 73$~eV. A powerlaw spectral component is not required by the data and contributes at most $\sim20\%$ to the total unabsorbed 0.5--10 keV flux of $\sim 9 \times 10^{-14}~\flux$. Such a soft quiescent spectrum is unusual for neutron stars with relatively high inferred magnetic fields and casts a different light on the interpretation of the hard spectral component, which is often attributed to magnetic field effects. For a distance of $5.5$~kpc, the estimated quiescent thermal bolometric luminosity is $\sim 6\times10^{32}~\lum$. If the thermal emission is interpreted as cooling of the neutron star, the observed luminosity requires that the system is quiescent for at least $\sim100$ years. Alternatively, enhanced neutrino emissions can cool the neutron star to the observed quiescent luminosity. 
\end{abstract}


\section{Introduction}\label{sec:intro}
Low-mass X-ray binaries (LMXBs) are binary star systems in which a neutron star or a black hole accretes matter from a (sub-) solar companion. Two phenomena are thought to uniquely identify the compact primary as a neutron star: coherent X-ray pulsations and type-I X-ray bursts. The latter are intense flashes of X-ray emission caused by thermonuclear runaway of the accreted matter on the surface of the neutron star. X-ray pulsations can be observed when the neutron star magnetic field is strong enough to funnel the accretion flow to the magnetic poles. 

Many neutron star LMXBs are transient and spend most of their lifetime in a quiescent state, during which they are dim with typical 0.5--10 keV X-ray luminosities of $L_q \sim 10^{31-33}~\lum$ \citep[e.g.,][]{heinke2009}. However, occasionally they exhibit outbursts during which their X-ray luminosity increases orders of magnitude to $L_X \sim 10^{36-38}~\lum$ \citep[2--10 keV; e.g.,][]{chen97}. The enhanced activity is ascribed to a sudden strong increase in the mass-accretion rate onto the neutron star, whereas little or no matter is accreted during quiescent episodes.

The spectra of quiescent neutron star LMXBs can typically be fitted with a soft thermal model, a hard powerlaw shape, or a combination of both. The hard spectral component has been attributed to non-thermal emission processes related to the magnetic field of the neutron star, e.g., accretion onto the magnetosphere or a pulsar wind mechanism \citep[e.g.,][]{campana1998}. 

The soft thermal component is generally interpreted as thermal emission from the neutron star surface. During accretion outbursts, a chain of nuclear reactions release heat deep in the neutron star crust \citep[e.g.,][]{haensel2008}, which is re-radiated during quiescent episodes. This results in an incandescent luminosity that is set by the long-term averaged mass-accretion rate of the system and the efficiency of neutrino emission processes occurring in the neutron star core \citep[e.g.,][]{brown1998}. Residual accretion onto the neutron star surface offers an alternative explanation for the quiescent thermal emission \citep[][]{zampieri1995}.

Galactic globular clusters are rich targets for studies of X-ray binaries, which can form in dynamical interactions in the dense cluster environments \citep[e.g.,][]{pooley2003}. Numerous low-luminosity X-ray sources have been found in globular clusters, amongst which are several candidate quiescent LMXBs \citep[e.g.,][]{verbunt1995_rosat, grindlay2001,heinke2006_terzan5}. Indeed, a few transient LMXBs have been identified in globular clusters during outburst episodes \citep[see e.g., the review by][]{verbunt2006}.

\begin{figure}
 \begin{center}
\includegraphics[width=8.0cm]{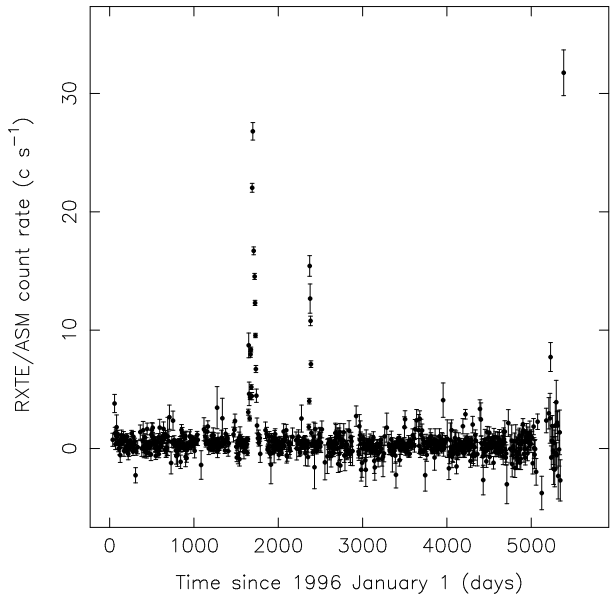}
    \end{center}
\caption[]{{\rxte/ASM 5-day averaged lightcurve of Terzan 5, showing activity from the globular cluster in 2000, 2003 and 2010.}}
 \label{fig:asm}
\end{figure}

\subsection{Terzan 5}\label{subsec:terzan5}
The globular cluster Terzan 5 has long been known to harbour at least one transient neutron star LMXB. In 1980, a number of X-ray bursts were observed with the {\it Hakucho} satellite \citep[][]{makishima1981}. Activity from Terzan 5 was also detected in 1984 with \exosat\ \citep[][]{warwick1988}, and in 1990/1991 with \rosat\ \citep[][]{verbunt1995_rosat,johnston1995}. Furthermore, \rxte/ASM observations have revealed three distinct X-ray outbursts since 1996: in 2000 \citep[][]{markwardt2000,heinke2003}, 2002 \citep[][]{wijnands2002_terzan5} and 2010 (see Figure~\ref{fig:asm}). 

The limited angular resolution of these instruments precludes pinpointing the X-ray source that causes the outbursts. However, \chan\ observations obtained during the 2000 outburst of Terzan 5 allowed for an accurate localization of the transient source that was active at that time \citep[][]{heinke2003}. Multiple X-ray bursts were observed and the 2000 transient was therefore associated with the X-ray burster that was detected by earlier X-ray missions (\hakuchoname/\exo). 
The quiescent counterpart of this transient neutron star LMXB has an unusually hard spectrum, with no clear evidence for the presence of a thermal emission component \citep[][]{wijnands2005}. A \chan\ study revealed $50$ distinct X-ray point sources within the half-mass radius of the cluster, several of which were proposed to be quiescent LMXBs \citep[][]{heinke2006_terzan5}.

Renewed activity from Terzan 5 was detected in 2010 October during \inte\ bulge scan monitoring observations \citep[][]{bordas2010}. Subsequent pointed \rxte/PCA observations detected coherent 11 Hz pulsations and type-I X-ray bursts, establishing the nature of the transient X-ray source as an accreting neutron star \citep[][]{strohmayer2010}. Timing studies of the X-ray pulsations provided a determination of the binary orbital period ($P_{\mathrm{orb}} = 21.27$~h), and the identification of the mass donor as a $\sim 0.4 - 1.5 ~\Msun$ main sequence or slightly evolved star \citep[][]{papitto2010}.

An accurate \swift\ localization revealed that the source active in 2010 was likely a different transient than the one detected in 2000 \citep[][]{kennea2010}. \chan\ observations confirmed this, thereby establishing the existence of a second transient neutron star LMXB in Terzan 5, named \source\ \citep[][]{pooley2010}. The angular separation between the two transients is only $\sim5.5''$ (see Figure~\ref{fig:ds9}), and due to the limited angular resolution of older X-ray missions it is unclear which of the two was responsible for other active episodes of Terzan 5. The 2010 transient corresponds to the X-ray source CX25 from the \chan\ study of \citet{heinke2006_terzan5}, which was marked as a candidate quiescent LMXB by these authors.

\section{Observations, data analysis and results}\label{sec:results}
\begin{figure}
 \begin{center}
\includegraphics[width=8.0cm]{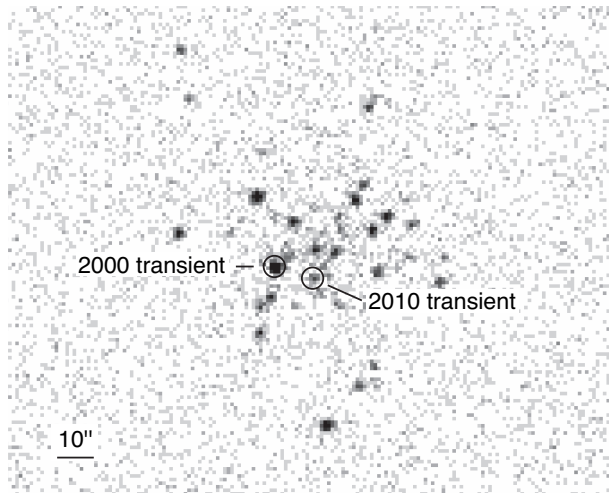}
    \end{center}
\caption[]{{\chan/ACIS image of the core of Terzan 5, obtained on 2009 July 15--16. The quiescent counterparts of both transient neutron star LMXBs that have been identified in this globular cluster are indicated.}}
 \label{fig:ds9}
\end{figure}

We searched the \chan\ data archive and found two ACIS-S observations of Terzan 5 in which no bright transient sources are active. This data is thus suitable to examine the quiescent properties of the newly detected 11 Hz X-ray pulsar. The first observation was performed on 2003 July 13--14 from 13:33--00:46 \textsc{ut} (ID 3798) for an exposure time of 39.5 ks \citep[see also][]{wijnands2005,heinke2006_terzan5}. The second observation had a duration of 36.4 ks and was obtained on 2009 July 15--16 from 17:21--04:14 \textsc{ut} (ID 10059). Both observations were carried out in the faint data mode, with the nominal frame time of 3.2~s and the globular cluster positioned on the S3 chip. Figure~\ref{fig:ds9} displays an image of the 2009 \chan\ observation.

Data reduction was carried out using the \textsc{ciao} software tools (v. 4.2). We reprocessed the level-1 data products with the task \textsc{acis$\_$process$\_$events}. The 2003 observation contained episodes of background flaring, which were excluded from further analysis, resulting in a net exposure time of 34.2~ks. We extracted source count rates and lightcurves using the tool \textsc{dmextract}, employing a $1''$ circular region centred at the coordinates reported by \citet{pooley2010}. Background events were collected from a circular region with a radius of $40''$, positioned on a source-free part of the CCD that was located $\sim1.4'$ West of the cluster core. 

The 2010 transient is detected at count rates of $(0.89\pm0.17)\times10^{-3}$ and $(1.11\pm0.18)\times10^{-3}~\cnts$ in the 2003 and 2009 data, respectively. A total of 30--40 net source photons were collected for both observations. Source and background spectra were obtained using \textsc{psextract}. We subsequently generated redistribution matrices (rmf) and ancillary response files (arf) with the tasks \textsc{mkacisrmf} and \textsc{mkarf}, respectively. The spectra were grouped using \textsc{grppha} to contain a minimum of 5 photons per bin, and fit within \textsc{Xspec} (v. 12.6) in the 0.5--8 keV energy range. Given the low number of photons, we also fitted the un-binned spectra using the W-statistic in \textsc{Xspec}. This yielded similar results as obtained for the binned spectra with the $\chi^2$ method applied, which indicates that the spectral fits are not biased by the small number of counts per bin. 

\begin{figure}
 \begin{center}
\includegraphics[width=8.0cm]{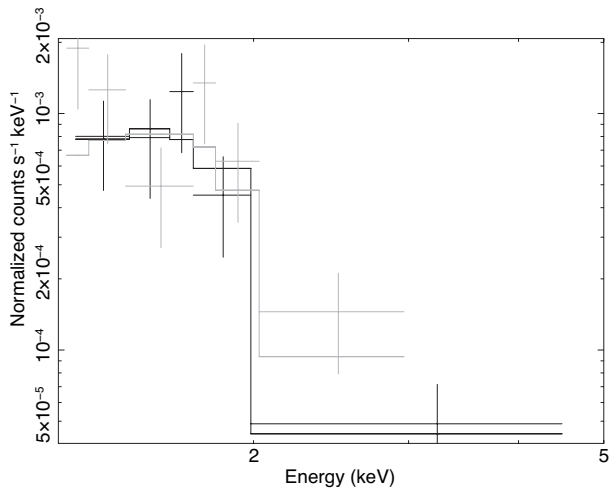}
    \end{center}
\caption[]{{\chan/ACIS quiescent spectra of the 11 Hz X-ray pulsar in Terzan 5 from 2003 (black) and 2009 (grey) data, along with a neutron star atmosphere model fit (solid lines).}}
 \label{fig:spec}
\end{figure}

Figure~\ref{fig:spec} displays the quiescent spectra of the new Terzan 5 transient. In both observations the spectrum is soft, with most photons detected below $\sim2$~keV. When fitting the 2003 and 2009 data separately, we found no spectral differences between the two observations. Therefore, we tied all spectral parameters between the two data sets. We assume a distance of $D=5.5$~kpc \citep[][]{ortolani2007} throughout this work and all quoted errors refer to $90\%$ confidence intervals. To account for the interstellar hydrogen absorption, we employ the \textsc{phabs} model with the default \textsc{Xspec} abundances and cross-sections.

Fitting the spectral data with a simple absorbed powerlaw yields $N_H= 1.3^{+3.2}_{-1.3} \times10^{22}~\mathrm{cm}^{-2}$, $\Gamma=4.8^{+4.9}_{-2.4}$ and $\chi_{\nu}=0.89$ for 8 degrees of freedom (dof). Such a large spectral index is usually interpreted as an indication that the spectrum has a thermal shape. We therefore tried an absorbed neutron star atmosphere model \textsc{nsatmos} \citep{heinke2006}, which is often used to fit the quiescent X-ray spectra of neutron stars. 

For the \textsc{nsatmos} model, we fix the neutron star mass and radius to canonical values of $M_{\mathrm{NS}}=1.4~\Msun$ and $R_{\mathrm{NS}}=10$~km, and the distance is frozen at $D=5.5$~kpc. Furthermore, we keep the normalization fixed at the recommended value of 1, which corresponds to the entire neutron star surface emitting. As such, the only free fit parameters are the hydrogen column density and the neutron star effective temperature. 
This yields $N_H=(2.1\pm0.9)\times10^{22}~\mathrm{cm}^{-2}$ and $kT^{\infty}=72.7\pm7.6$~eV ($\chi_{\nu}=0.92$ for 9 dof). The corresponding 0.5--10 keV unabsorbed flux is $(9.4\pm 4.6)\times10^{-14}~\flux$. For a distance of 5.5 kpc, the associated luminosity is $(3.4\pm 1.7)\times10^{32}~\lum$. Extrapolating the model fit to the 0.01--100 keV energy range yields an estimate of the thermal bolometric flux of $(1.7\pm 0.7)\times10^{-13}~\flux$. This implies a thermal bolometric luminosity of $L_q = (6.2\pm2.5)\times10^{32}~(D/5.5~\mathrm{kpc})^2~\lum$. The \textsc{nsatmos} model fit is plotted in Figure~\ref{fig:spec}.

Although the data can be adequately fit by a neutron star atmosphere model alone, we explored fits with an additional powerlaw to probe the contribution of a possible non-thermal emission component. We fix the spectral index to $\Gamma=1.5$, which is a typical value observed for neutron star transients in quiescence. For the two-component model we obtain $N_H=(1.6\pm1.6)\times10^{22}~\mathrm{cm}^{-2}$ and $kT^{\infty}=67.1\pm10.5$~eV ($\chi_{\nu}=0.97$ for 8 dof). The powerlaw is not significant, but may contribute $6^{+13}_{-6}\%$ to the total unabsorbed 0.5--10 keV luminosity of $2.5\pm 2.5 \times10^{32}~(D/5.5~\mathrm{kpc})^2~\lum$. The inferred thermal bolometric luminosity is $4.6\pm 3.7 \times10^{32}~(D/5.5~\mathrm{kpc})^2~\lum$. Since the additional powerlaw component is not required to fit the data, we consider the calculated fractional contribution as an upper limit. Choosing different spectral indices ($\Gamma=1-2$) yields spectral parameters and fluxes that are consistent within the errors with the results for $\Gamma=1.5$. 

The values obtained for the hydrogen column density are comparable to the results of \citet{heinke2006_terzan5}, who performed spectral fits for the 16 brightest X-ray sources in Terzan 5 (which does not include the new 2010 transient), and found an average value of $N_H\sim1.9 \times10^{22}~\mathrm{cm}^{-2}$. 
However, \citet{bozzo2010} report $N_H=(5.0\pm0.8)\times10^{21}~\mathrm{cm}^{-2}$, based on spectral analysis of \swift/XRT data obtained during the 2010 outburst. Therefore, we have also performed fits to the quiescent data with the hydrogen column density fixed at $N_H=5.0\times10^{21}~\mathrm{cm}^{-2}$. For this value, the thermal bolometric luminosity decreases by a factor of $\sim2.5$ and the allowed fractional powerlaw contribution increases to $\sim25^{+26}_{-15}\%$.

\section{Discussion}\label{sec:discuss}
We report on the spectral analysis of the newly discovered 11 Hz X-ray pulsar \source\ (\intename) in the globular cluster Terzan 5 during quiescence. Using two archival \chan\ observations, carried out in 2003 and 2009, we show that the quiescent spectrum is dominated by thermal emission that fits to a neutron star atmosphere model with a temperature of $kT^{\infty} \sim 73$~eV. The inferred thermal bolometric luminosity is $L_q \sim 6 \times 10^{32}~(D/5.5~\mathrm{kpc})^2~\flux$. If a powerlaw is included in the fits, this spectral component contributes $\lesssim20\%$ to the total unabsorbed 0.5--10 keV luminosity of $\sim 3 \times 10^{32}~(D/5.5~\mathrm{kpc})^2~\flux$. There is no evidence for spectral variations in the quiescent emission between the 2003 and 2009 observations. 

\subsection{The powerlaw spectral component}\label{subsec:powerlaw}
It is interesting to compare the quiescent spectral properties of the 2010 Terzan 5 transient with results obtained for the accreting millisecond X-ray pulsars (AMXPs). The quiescent spectra of five AMXPs could be studied in detail: \ngcbron, Aql X-1, \sax, \xtepulsar\ and \igrpulsar. The former two are both intermittent X-ray pulsars (i.e., the pulsations are detected only sporadically during outburst), that have predominantly soft quiescent X-ray spectra and a 0.5--10 keV luminosity of $\sim 10^{33}~\lum$ \citep[][]{rutledge2001,cackett2005}. \sax, \xtepulsar\ and \igrpulsar, have quiescent 0.5--10 keV luminosities on the order of $\sim 10^{32}~\lum$. Of the three, only \igrpulsar\ shows evidence for thermal emission, contributing $\sim 40\%$ to the total 0.5--10 keV flux \citep[][]{heinke2009}, while the quiescent spectra of the other two are completely dominated by a hard spectral component \citep[][]{campana2002,wijnands05_amxps}. 

Our limited understanding of the origin of hard quiescent X-ray emission makes it difficult to interpret the difference between the 2010 Terzan 5 transient and other X-ray pulsars. However, if the powerlaw spectral component is related to the magnetic field of the neutron star \citep[][]{campana1998}, we would have expected to detect a hard X-ray spectrum for the 11 Hz pulsar in Terzan 5, since this source must have a substantial magnetic field \citep[][Cavecchi et al. in prep.]{papitto2010}. A possible explanation for the soft X-ray spectrum of the 2010 Terzan 5 transient is that little matter is available in quiescence to interact with the neutron star magnetic field, so that the powerlaw emission component that may arise from this process is strongly reduced. The effects of the strong magnetic field on the neutron star spectrum might also play a role \citep[e.g.,][]{zavlin2002}.

\subsection{The quiescent thermal emission}\label{subsec:thermal}
The observed thermal quiescent luminosity of the 11 Hz pulsar in Terzan 5 can be used to estimate the duty cycle of the system, by equating this value to the luminosity that is expected to be radiated due to heating of the neutron star. 
Assuming that the observed accretion luminosity provides a measure for the amount of matter that is accreted onto the neutron star, we can deduce the mass-accretion rate during outburst from the average outburst flux. 
\citet{papitto2010} infer a 0.1--100 keV unabsorbed flux of $\sim 1 \times 10^{-8}~\flux$ from fitting \rxte/PCA and HEXTE spectral data obtained during the 2010 outburst of Terzan 5. This yields an estimated bolometric accretion luminosity of $L_{acc} \sim 4\times10^{37}~(D/5.5~\mathrm{kpc})^2~\lum$.

For an accretion luminosity given by $L_{acc} = (G M_{NS}/R_{NS}) \langle \dot{M}_{ob} \rangle$, the average accretion rate during outburst is $\langle \dot{M}_{ob} \rangle \sim 2 \times 10^{17}~\mathrm{g~s}^{-1} \sim 3\times10^{-9}~\mdot$, for canonical neutron star parameters of $M_{NS} = 1.4~\Msun$ and $R_{NS} = 10$~km. The nuclear reactions induced by the accretion of matter are expected to result in a quiescent bolometric luminosity that is given by $L_q = \langle \dot{M} \rangle Q_{nuc} / m_u$ \citep[e.g.,][]{brown1998,colpi2001}. Here, $Q_{nuc} \sim 2$ MeV is nuclear energy deposited in the crust per accreted baryon \citep[e.g.,][]{haensel2008}, $m_u=1.66\times10^{-24}$~g is the atomic mass unit and $\langle \dot{M} \rangle = \langle \dot{M}_{ob} \rangle \times t_{ob} / t_{rec}$ is the time-averaged mass-accretion rate of the system. The ratio of the outburst duration ($t_{ob}$) and the recurrence time ($t_{rec}$) represents the duty cycle.

Using the above equation, the observed quiescent thermal bolometric luminosity of $L_q \sim 6 \times 10^{32}~(D/5.5~\mathrm{kpc})^2~\lum$ suggests a time-averaged mass-accretion rate of $\langle \dot{M} \rangle \sim 5\times10^{-12}~\mdot$. Combined with the estimated mass-accretion rate during the 2010 outburst, this would imply that the system must have a duty cycle on the order of $\sim 0.1 \%$, provided that standard cooling mechanisms are operating in the core. The 2010 outburst commenced around October 10 and is ongoing during \rxte/PCA pointed observations performed on November 19.\footnote{Terzan 5 is unobservable with \rxte\ between 2010 November 19 and  2011 January 17 due to Sun-angle constraints, nor can it be observed with other X-ray satellites during that epoch.} The outburst thus has a duration of  $>6$~weeks, so the estimated duty cycle would imply that the source must spend $\gtrsim100$~yr in quiescence. This increases further if the outburst is observed to continue for a longer time. 

Although it is plausible that the 2010 Terzan 5 transient has a very low duty cycle, an alternative explanation is that enhanced neutrino emission mechanisms are operating in the core, which cool the neutron star down to the observed quiescent luminosity. Within our current understanding, this would suggest that the neutron star in this X-ray binary is relatively massive \citep[e.g.,][]{colpi2001}, or that the core composition includes different forms of matter \citep[e.g., pions or kaons;][]{yakovlev2004}. It is not obvious that the neutron star in this transient LMXB would be particularly massive, since both its high inferred magnetic field \citep[$B \sim10^{9}-10^{10}$~G or possibly even higher;][Cavecchi et al. in prep.]{papitto2010} and relatively slow spin period ($P_s = 11$~Hz) are suggestive of a relatively young system. 

From modelling the quiescent spectral data we find a neutron star effective temperature, as observed by a distant observer, of $kT^{\infty}\sim73$~eV. In the neutron star frame this translates into a value of $kT\sim95$~eV ($T \sim 1 \times 10^{6}$~K) for $M_{\mathrm{NS}}=1.4~\Msun$ and $R_{\mathrm{NS}}=10$~km (corresponding to a gravitational redshift of $1+z=1.3$). Using the model calculations of \citet{brown08} we can obtain a rough estimate of the neutron star core temperature of $\sim3\times10^{7}$~K. For standard (i.e., slow) core neutrino emission processes, such a core temperature yields a neutrino luminosity on the order of $L_{\nu} \sim 10^{28}~\lum$ \citep[cf. figure 2 of][]{schaab1999}. This is negligible compared to the observed photon emissions of $L_q \sim 6 \times 10^{32}~(D/5.5~\mathrm{kpc})^2~\lum$. However, for a sample enhanced cooling model (direct Urca), the neutrino luminosity is on the order of $L_{\nu} \sim10^{35}~\lum$ \citep[cf. figure 2 of][]{schaab1999}, and thus the dominant cooling mechanism. For such neutrino losses a duty cycle of roughly $\sim25\%$ is required to explain the observed quiescent thermal luminosity. The time-averaged mass-accretion rate for this scenario is $\langle \dot{M} \rangle \sim 8\times10^{-10}~\mdot$.

The outburst duration of the new Terzan 5 transient is unconstrained, but for a typical value of $\sim 2-12$~months, a duty cycle of $\sim25\%$ implies a recurrence time of $\sim1-4$~yr. This would suggest that some of the previous outbursts from Terzan 5 might have originated from the 11 Hz X-ray pulsar. \chan\ observations leave no doubt that the 2000 outburst of Terzan 5 was caused by the other transient LMXB located in this cluster \citep[cf.][]{heinke2003,pooley2010}. However, lack of high-spatial resolution observations during previous active periods of Terzan 5 (i.e., 1980, 1984, 1990, 1991 and 2002; see Section~\ref{sec:intro}) do not rule out this possibility. Detailed studies of the X-ray burst behaviour can potentially shed more light on this. 

We note that the non-detection of quiescent thermal emission for the 2000 Terzan 5 transient led \citet{wijnands2005} to conclude that this source must either spend hundreds of years in quiescence, or be subject to enhanced neutrino cooling. If both Terzan 5 LMXBs undergo slow core cooling and spend long episodes in quiescence, the other outbursts observed from this globular cluster must have been caused by another transient source. There are several candidate quiescent LMXBs identified in Terzan 5 \citep[][]{heinke2006_terzan5}. Alternatively, at least one of the neutron star LMXBs undergoes enhanced neutrino core cooling.

When inferring the quiescent bolometric luminosity of the 2010 Terzan 5 transient we assumed a distance of $D=5.5$~kpc, as inferred by \citet{ortolani2007}. We note that \citet{cohn2002} report a distance towards Terzan 5 of $D=8.7$~kpc, which would increase the thermal bolometric luminosity inferred in this work by a factor of $\sim2$. However, for a larger distance the inferred accretion luminosity is also increased, so in practise this does not affect the estimated duty cycle. Furthermore, as discussed in \citet{ortolani2007}, $D=5.5$~kpc can be considered a more reliable distance estimate. 
We note that if the observed quiescent thermal emission component is due to residual accretion onto the neutron star surface \citep[][]{zampieri1995}, the interior temperature of the neutron star must be lower than we infer here, thus requiring a smaller duty cycle. 

The new transient source discovered in Terzan 5 became very bright during its 2010 outburst, reaching up to $L_X \sim 10^{38}~\lum$ \citep[][]{altamirano2010_2}. Combined with the relatively low inferred neutron star temperature during quiescence, this makes the source a potential target to search for cooling of the neutron star crust, once the accretion ceases. \citet{brown1998} argue that the neutron star crust can become significantly heated for systems in which the outburst luminosity is much higher than the quiescent level. Once the accretion ceases, the thermal relaxation of the neutron star crust might become observable as a gradual decrease in neutron star effective temperature. \\

\noindent {\bf Acknowledgements.}\\
This work was supported by the Netherlands Research School for Astronomy (NOVA) and made use of the \chan\ public data archive. RW acknowledges support from a European Research Council (ERC) starting grant. The authors are grateful to the anonymous referee for providing thoughtful comments that helped improve this manuscript.

\label{lastpage}
\end{document}